\documentclass[journal = jacsat]{achemso}
\setkeys{acs}{articletitle = false}

\usepackage{hyperref, graphicx, amsmath, amssymb, xcolor}
\graphicspath{{pictures/}} 
\usepackage{subcaption} 
\usepackage{caption} 
\usepackage{wrapfig}
\usepackage{float}
\usepackage{mathtools}
\usepackage{indentfirst}

 \author{Elena Titova}
 \affiliation{Center for Photonics and 2D Materials, Moscow Institute of Physics and Technology, Dolgoprudny, 141700, Russian Federation}
\alsoaffiliation{Programmable Functional Materials Lab, Brain and Consciousness Research Center, Moscow, 121205, Russia}
 \email{titova.elenet@gmail.com}
 
 \author{Dmitry Mylnikov}
 \affiliation{Center for Photonics and 2D Materials, Moscow Institute of Physics and Technology, Dolgoprudny, 141700, Russian Federation}
 
  \author{Mikhail Kashchenko}
 \affiliation{Center for Photonics and 2D Materials, Moscow Institute of Physics and Technology, Dolgoprudny, 141700, Russian Federation}
\alsoaffiliation{Programmable Functional Materials Lab, Brain and Consciousness Research Center, Moscow, 121205, Russia}
 
 \author{Sergey Zhukov}
 \affiliation{Center for Photonics and 2D Materials, Moscow Institute of Physics and Technology, Dolgoprudny, 141700, Russian Federation}

\author{Kirill Dzhikirba}
\affiliation{Institute of Solid State Physics, RAS, Chernogolovka 142432, Russian Federation}
 
\author{Kostya Novoselov}
\affiliation{Institute for Functional Intelligent Materials, National University of Singapore, Singapore, 117575, Singapore}
\alsoaffiliation{Programmable Functional Materials Lab, Brain and Consciousness Research Center, Moscow, 121205, Russia}
 
 \author{Denis Bandurin}
\affiliation{Department of Materials Science and Engineering, National University of Singapore, 117575 Singapore}

 \author{Georgy Alymov}
 \affiliation{Center for Photonics and 2D Materials, Moscow Institute of Physics and Technology, Dolgoprudny, 141700, Russian Federation}

 \author{Dmitry Svintsov}
 \affiliation{Center for Photonics and 2D Materials, Moscow Institute of Physics and Technology, Dolgoprudny, 141700, Russian Federation}

\title{Ultralow-noise terahertz detection by p-n junctions in gapped bilayer graphene}

\begin{document}

\maketitle
\date{\today}

\begin{abstract}

Graphene shows a strong promise for detection of terahertz (THz) radiation due to its high carrier mobility, compatibility with on-chip waveguides and transistors, and small heat capacitance. At the same time, weak reaction of graphene's physical properties on the detected radiation can be traced down to the absence of band gap. Here, we study the effect of electrically-induced band gap on THz detection in graphene bilayer with split-gate $p$-$n$ junction. We show that gap induction leads to simultaneous increase in current and voltage responsivities. At operating temperatures of $\sim 25$~K, the responsivity at 20~meV band gap is from 3 to 20 times larger than that in the gapless state. The maximum voltage responsivity of our devices at $0.13$~THz illumination exceeds 50~kV/W, while the noise equivalent power falls down to 36~fW/Hz$^{1/2}$. These values set new records for semiconductor-based cryogenic terahertz detectors, and pave the way for efficient and fast terahertz detection.

\end{abstract}

keywords: bilayer graphene, photodetectors, terahertz

Terahertz (THz) radiation finds potential applications in multiple areas of technology, such as high-speed wireless communications~\cite{Nagatsuma2016,Sengupta2018}, medical imaging~\cite{Yu2019}, security scanning~\cite{Tzydynzhapov2020}, and defect inspection~\cite{Ellrich2020}. The complexity of detecting terahertz radiation stems from its is weak absorption by most semiconductor materials, as it does not induce interband transitions. The detectors using radio-engineering rectification principles, in turn, suffer from large $RC$-delays at terahertz frequencies. These complexities can be largely overcome when using graphene as a material for terahertz detection, due to its high electron mobility, gate-tunable carrier density, almost frequency-independent light absorption, and small heat capacitance ~\cite{bonaccorso_graphene_2010}. Owing to its atomically thin body, graphene is compatible with silicon technology~\cite{miseikis_ultrafast_2020,goossens_broadband_2017,brenneis_thz-circuits_2016} and thus well suited for on-chip integrated optoelectronics~\cite{an_perspectives_2022, youngblood_integration_2016,vangelidis_unbiased_2022}. 

There have been numerous successful demonstrations of THz detectors with graphene as an active element, reviewed in~\cite{rogalski_graphene-based_2019}. Importantly, for zero-bias detection of THz radiation enabling low noise, one has to introduce structural asymmetry into the device. Examples of such asymmetry are dissimilar metals for source and drain contacts~\cite{cai_sensitive_2014}, feeding the radiation from antenna coupled between source and gate of the transistor~\cite{vicarelli_graphene_2012,bandurin_resonant_2018}, shift of gate electrode with respect to mid-channel~\cite{viti_thermoelectric_2020,shabanov_optimal_2021,Muravev2012e}, and complex asymmetric patterns of contacts~\cite{Auton2017e} and metallization~\cite{delgado-notario_asymmetric_2020,delgado-notario_enhanced_2022}. One of the most well-understood architectures for zero-bias radiation detection in graphene is the lateral $p$-$n$ junction. Compared to bulk semiconductors, this junction in 2d materials does not require chemical doping and can be introduced fully electrically. Electromagnetic detection by gate-induced $p$-$n$ junctions was demonstrated for monolayer graphene~\cite{castilla_fast_2019,asgari_terahertz_2022}, twisted bilayer~\cite{zhang_multimode_2022}, and suspended naturally-stacked bilayer graphene (BLG) in the gapless state~\cite{jung_microwave_2016}. All these devices demonstrated high responsivities and low noise-equivalent power ($\sim 80$ pW/Hz$^{1/2}$ at room temperature). 

Quite surprisingly, the effect of band gap opening in graphene bilayer $p$-$n$ junction on its THz detection performance has not been studied yet. At the same time, the gap was shown to considerably enhance the responsivity of gate-coupled transistor-based detectors with BLG channel~\cite{gayduchenko_tunnel_2021}. It may be expected that gap opening would result in emergence of new THz rectification pathways, particularly, via non-linearities of interband tunnel junctions between regions with dissimilar doping. Other rectification mechanisms well-studied for monolayer and gapless bilayer, can also be enhanced by the presence of the gap. In particular, photo-thermoelectric~\cite{gabor_hot_2011} and resistive self-mixing effects~\cite{Sakowicz2011} are both proportional to the sensitivity of conductance to the gate voltage, $d\ln G/d V_g$. As this quantity is maximized once the Fermi level is located within the gap, the photoresponse of BLG-based detectors may be favored by gap opening.

In this work, we study the sub-THz photoresponse in BLG-based detectors with gate-induced $p$-$n$ junction focusing on the gapped state of the channel. We find that induction of gap indeed results in enhanced responisivty, both for rectified current and rectified voltage. The observed sign and pattern of photovoltage with varying the carrier densities at cryogenic temperatures ($T\approx 25$~K) can be explained by the photo-thermoelectric scenario. At the same time, we evidence the enhancement of photovoltage as the doping under two top gates is opposite to that in near-contact regions. This feature may be attributed either to extra rectification mechanisms, like rectification by tunnel junctions at the contacts. Instructively, at nearly room temperature ($T \approx 300$~K), the overall sign of photoresponse changes. This indicates on the dominance of non-thermoelectric rectification mechanisms, presumably, resistive self-mixing.

\section*{Results and Discussion}
\begin{figure}[H]
    \centering
    \includegraphics[width=1\textwidth]{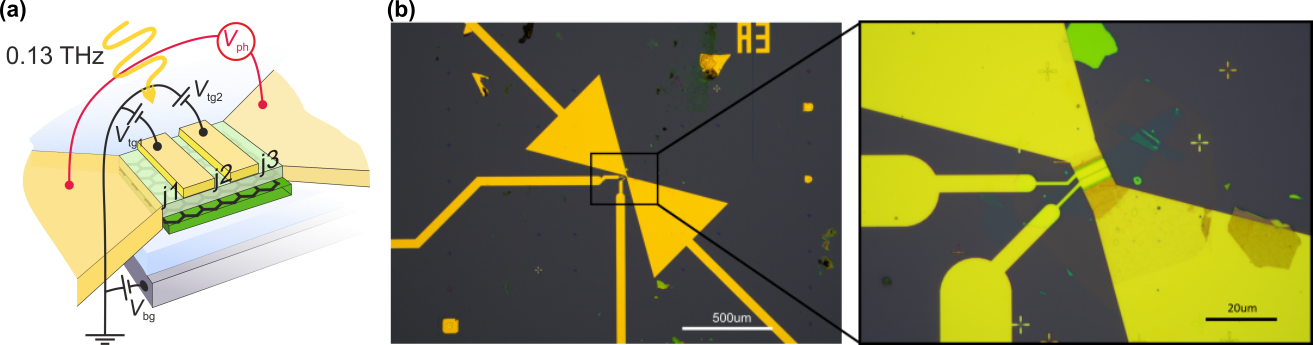}
    \caption{{\bf Split-gate graphene bilayer THz detector.} (a) Schematic of the device: source (S) and drain (D) are connected to antenna arms, the photovoltage $V_{\rm ph}$ is developed between the same contacts. Top gates (TG$_1$ and TG$_2$) control the carrier densities at both sides, while back gate (BG) induces the global band gap (b) Micro-photographs of sample A under different magnification}
    \label{fig:structures}
\end{figure}

\textbf{Split-gate graphene bilayer transistors.} Our devices are based on encapsulated bilayer graphene channel with split top gate and global bottom gate, shown in Fig.~\ref{fig:structures}. The fabrication of device is described in the Methods section. The tri-gate architecture enables us to simultaneously induce the band gap in the channel and control the carrier density under each of top gates independently. We feed the radiation from IMPATT-diode source with frequency $f=130$ GHz by illuminating the bow-tie antenna connected between source and drain of the device. The emerging rectified voltage builds up between the source (grounded) and  the drain. The latter is connected to the lock-in amplifier. We have studied two samples, marked as A and B, with parameters listed in Table~\ref{tbl:samples}. Sample A demonstrated higher responsivity, potentially due to better antenna design; yet it showed weaker effect of back gate voltage on resistance, probably due to unintentional built-in impurities. 

\begin{table}
  \caption{Parameters of the samples}
  \label{tbl:samples}
  \begin{tabular}{lll}
    \hline
    Parameter  & Sample A & Sample B \\
    \hline
    Intergate gap & 150 nm & 200 nm  \\
    Channel length  & 6 $\mu$m &  6 $\mu$m \\
    Channel width   & 9 $\mu$m & 3 $\mu$m \\
    Top gate width  &  2.4 $\mu$m  & 2.4 $\mu$m \\
    Top dielectric &  30 nm hBN  & 60 nm hBN  \\
    Bottom dielectric  & (280 +  50) nm SiO$_2$ + hBN & (280 + 40) nm SiO$_2$ + hBN\\
    Carrier mobility & 45000 cm$^2$/(V$\cdot$s) & 30000 cm$^2$/(V$\cdot$s) \\
    Responsivity  & 50.5 kV/W  & 1.8 kV/W \\
    \hline
  \end{tabular}
\end{table}

\begin{figure}[H]
    \centering
    \includegraphics[width=1\textwidth]{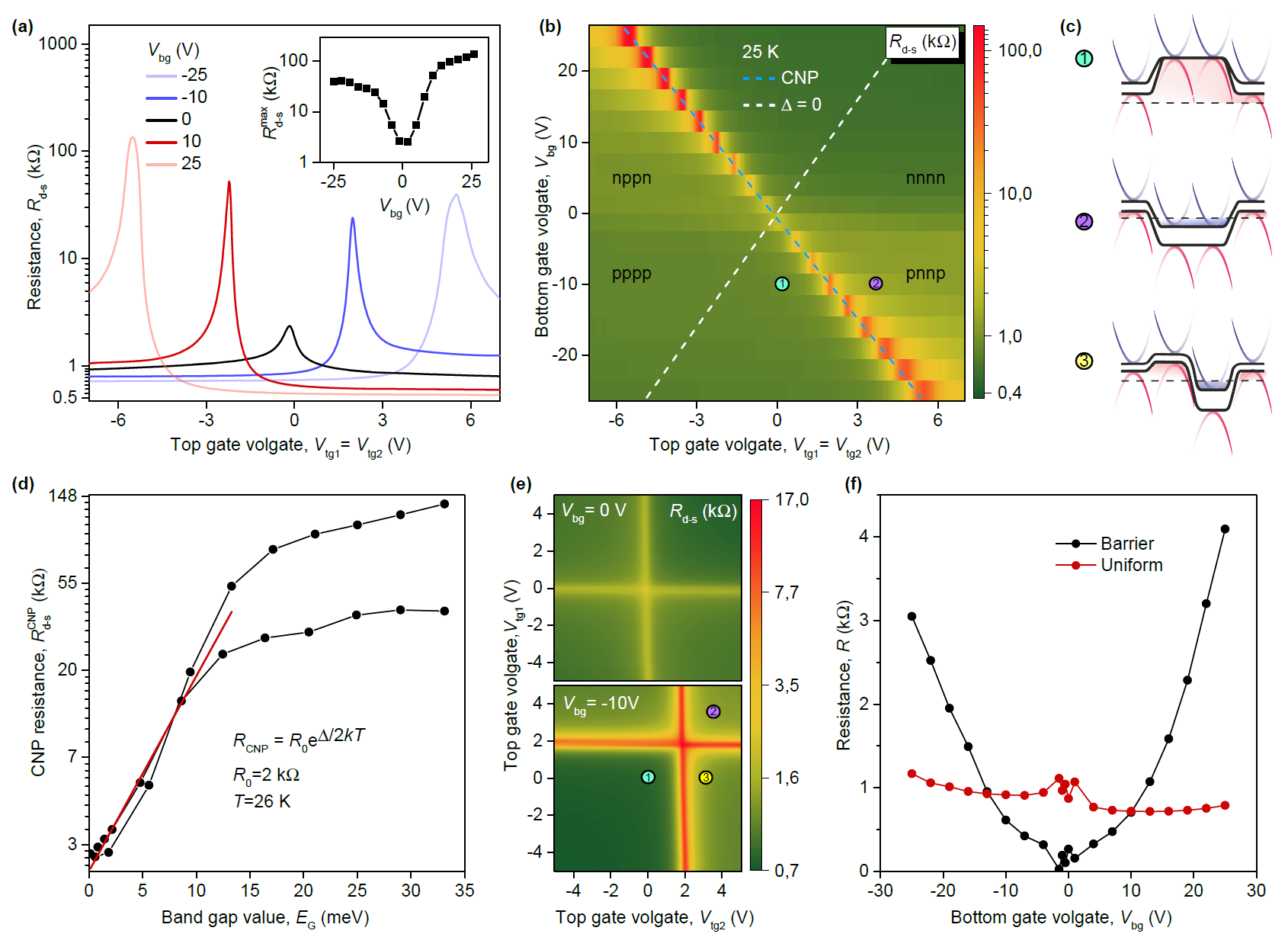}
    \caption{ {\bf Transport characteristic of BLG $p$-$n$ junction (sample B).} (a) Resistance $R_{SD}$ vs top gate voltage $V_{\rm tg1} = V_{\rm tg2}$ at a series of back gate voltages, from $-25$ V to $+25$ V, $T=25$ K. Inset: maximum resistance vs back gate voltage (b) Color map of resistance vs two gate voltages, $R(V_{\rm tg},V_{\rm bg})$ (c) Band diagrams of BLG FET for three characteristic points, marked by colored circles in (b,e). (d) Dependence of maximum resistance on band gap $E_G$ (black), computed as $E_G = \alpha |C_{\rm tg} V_{\rm tg} - C_{\rm bg} V_{\rm bg}|/\varepsilon_0$ along with the exponential fit (red) (e) Dependence of resistance on top gate voltages $R(V_{\rm tg 1},V_{\rm tg2})$ in the gapless (top) and gapped (bottom) states (f) Extracted resistances of barrier between $p$ and $n$ regions (black) and their bulk (red) vs back gate voltage.}
    \label{fig:resistivities}
\end{figure}

\textbf{Transport indicators of gap opening.} We start the discussion of our results by presenting the transport evidence of gate-induced band gap at cryogenic temperatures. Figure ~\ref{fig:resistivities} displays the drain-to-source graphene resistance ($R_{\rm SD}$) as a function of bottom $V_{bg}$ and top $V_{tg}$ gate voltages for sample B. Here, the voltages at the left and right top gates are identical, $V_{t1} = V_{t2} = V_{tg}$. In agreement with previous studies, the maximum resistance is realized at the map diagonal. Its equation reads as $C_{tg} V_{tg} + C_{bg} V_{bg} = 0$, where $C_{tg}$ and $C_{bg}$ are the specific capacitances between the channel and the respective gates. The other set of map diagonals, given by  $C_{tg} V_{tg} - C_{bg} V_{bg} = {\rm const}$, corresponds to the constant values of the induced band gap $E_G$. The absolute value of $E_G$ is maximized in the top left and bottom right corners of the map, which manifests itself in an abrupt, two order-of-magnitude resistance enhancement at the charge neutrality point (CNP). We further record the CNP resistance $R_{CNP}$ as a function of computed band gap $E_G = \alpha |C_{\rm tg} V_{\rm tg} - C_{\rm bg} V_{\rm bg}|/\varepsilon_0$, where $\alpha \approx 0.05$~eV$\cdot$nm/V~\cite{Falko_bandgap_theory,alymov_abrupt_2016}, Fig.~\ref{fig:resistivities} (d). These dependences are exponential, in agreement with Arrhenius law.

Another important indicator of band gap opening, scarcely discussed in the literature, is the emergence of a 'high shoulder' in the $R(V_{\rm tg})$-dependence at finite back gate voltages $V_{\rm bg}$ (Fig.~\ref{fig:resistivities} a). This enhanced resistance is also clearly seen in the resistance map, Fig.~\ref{fig:resistivities} b. It appears if the doping of the mid-channel, controlled by both gates, is opposite to that of near-contact regions, controlled by the bottom gate only. Such extra resistance can be thus attributed to the resistance of gate-induced junctions. In clean encapsulated samples, the transport through such junctions is dominated by elastic interband tunneling.

Contrary to the well-studied case of single-layer graphene~\cite{castilla_fast_2019}, the resistance of induced $p$-$n$ junction can make a considerable fraction of bulk resistance or even exceed it. To demonstrate that, we create this junction on-demand in the middle of the channel, by fixing the back gate voltage and sweeping TG$_1$ and TG$_2$ independently. The resulting maps of $R(V_{\rm tg1}, V_{\rm tg2})$ are shown in Fig.~\ref{fig:resistivities} (e). It's clearly seen that the resistance is minimized for fully uniform doping (quadrant 3 of the maps at $V_{\rm bg}<0$) and is increased once the junction is present in the channel. The junction resistance $R_J$ can be determined as $R_{\rm pppp} - R_{\rm pnnp}$, where the lower indices indicate the doping of four subsequent regions of the channel. The extracted junction resistance varies from zero up to $\sim 4$~k$\Omega$ at $V_{\rm bg} = \pm 25$~V, and $\sim 4$ times exceeds the bulk resistance. The latter is weakly sensitive to the gap $E_G$ and is rather determined by channel doping. 

\begin{figure}[H]
    \centering
    \includegraphics[width=1\textwidth]{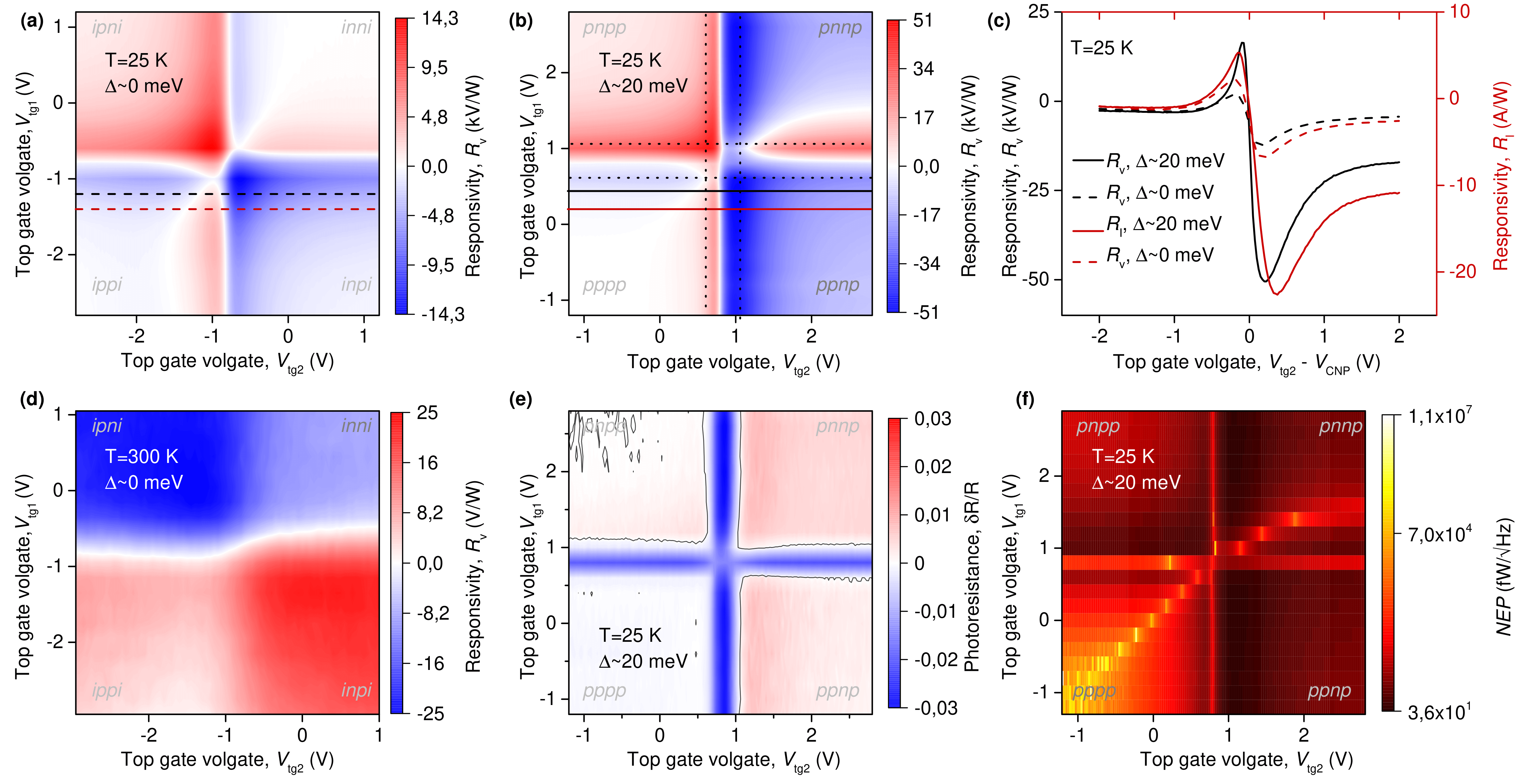}
    \caption{{\bf Sub-THz photoresponse of BLG $p$-$n$ junction.} (a,b) Voltage responsivity measured at $T=25$~K vs carrier densities at the two sides of junction, controlled by voltages $V_{\rm tg1}$ and $V_{\rm tg2}$, in the gapless state [$V_{\rm bg} = 3.6$~V, (a)] and in the gapped state [$V_{\rm bg} = -16.4$~V, (b)] (c) Comparison of current and voltage responsivities in the gapless (dashed lines) and gapped (solid lines) states along the horizontal line cuts in panels (a,b) (d) Density-dependent voltage responsivity at room temperature (e) Map of THz photoresistance of gapped BLG junction (f) Calculated noise-equivalent power of gapped BLG $p$-$n$ junction at $T=25$~K and $V_{\rm bg} = -16.4$~V}
    \label{fig:photoresults}
\end{figure}

\textbf{Photoresponse of gapped $p$-$n$ junction in graphene bilayer.} Having proved the strong impact of band gap and induced $p$-$n$ junctions on device resistivity, we proceed to discussion of THz detection by the structure. The experimentally measured zero-bias photovoltage maps for the sample A is presented in Fig.~\ref{fig:photoresults}. In these maps, we vary the doping at the two sides of junction (by sweeping TG$_1$ and TG$_2$-voltages) at approximately fixed band gap set by $V_{\rm bg}$. To keep up with practically relevant figures of merit, we recalculate the detector photovoltage $V_{\rm ph}$ into voltage responsivity $r_V =  V_{\rm ph}/P$, where $P$ is the radiation power reaching the device (see Methods for calibration procedure).

In agreement with general intuition, the responsivity of our BLG-based detectors is maximized once the sides of the junction are doped oppositely, quadrants II and IV of the maps. The average photosignal for identically-doped sides, quadrants I and III, is always lower than for oppositely-doped ones, which implies that the mid-channel junction plays the major role in the photoresponse. At identical doping polarities, diagonal of quadrants I and III, the photovoltage is absent, consonant with symmetry considerations. The photovoltage changes sign six times along the closed loop in the $r_V(V_{\rm tg1},V_{\rm tg2})$ map. 

Induction of band gap results in overall enhancement of $r_V$ without any changes in the symmetry of the maps. This is clearly seen by comparing the panels (a) and (b) of Fig.~\ref{fig:photoresults}, recorded at back gate voltages $V_{\rm bg} = 3.6$~V (charge neutrality point) and $V_{\rm bg} = -16.4$~V. For sample B, the maximum responsivity enhancement factor reaches 20, while for sample A it is up to 3. The {\it voltage} responsivity enhancement with induction of the gap is partially due to the enhancement of BLG dc resistance. Still, this trivial effect does not full explain the variations of photoresponse. Indeed, the current responsivity $r_I = I_{\rm ph}/P$, being less sensitive to the load resistance, also demonstrates a strong increase with band gap induction [Fig.~\ref{fig:photoresults} (c)].

We complete our empirical analysis of photovoltage at cryogenic temperatures by correlating the positions of responsivity maxima on $r_V(V_{\rm tg1},V_{\rm tg2})$-maps and the maps of photoresistance $\Delta R_{\rm SD}$. The latter was measured simultaneously with photoresponse [Fig.~\ref{fig:photoresults} (e)] and discussed in detail in Ref.~\cite{Our_photoconductivity}. Such a comparison reveals that $r_V$ maxima are located in the vicinity of photoresistance zeros. In the most common bolometric scenario, $\Delta R_{SD}$ turns to zero as the Fermi level crosses the bottom of conduction (or top of the valence) band, switching BLG between metallic and semiconductor states. Thus, the roots of enhanced photoresponse at increased gap should be searched for in some physics near the band edges.

Quite unexpectedly, the sign of photovoltage flips as the sample is heated from $T \sim 25$~K to the room temperature [see Fig.~\ref{fig:photoresults} (d)]. This flip occurs systematically both in samples A and B at all accessible band gaps, indicating on the change in the detection physics. At room temperature, the photovoltage varies only slightly (within 20~\%) with gap induction. This is consistent with the fact that magnitude of the gap is less than $2kT \approx 50$~meV at $T=300$~K.

\begin{figure}
    \centering
    \includegraphics[width=0.9\textwidth]{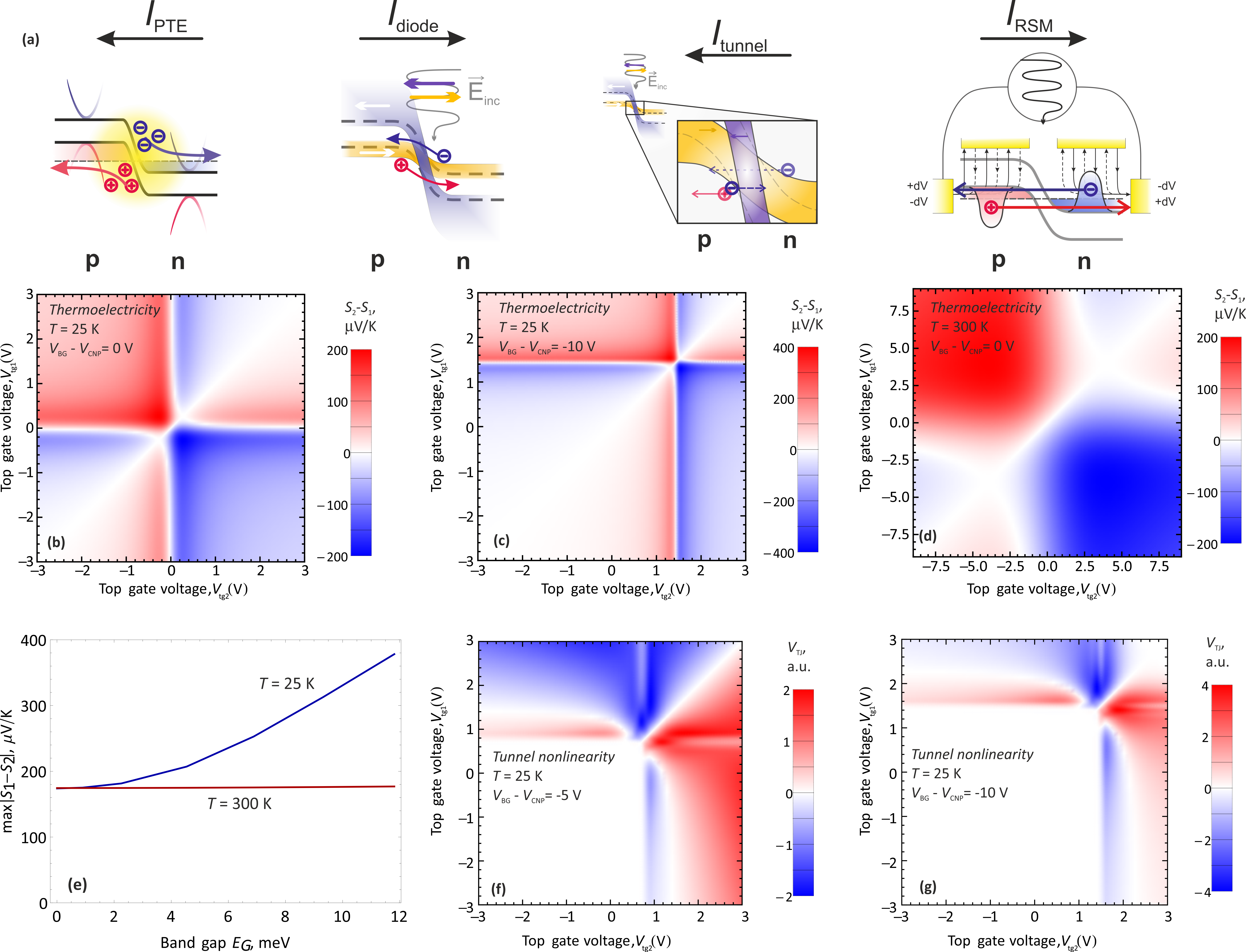}
    \caption{{\bf Modeling of photoresponse in gapped $p$-$n$ junction.} (a) Signs of photocurrent for conventional THz rectification mechanisms: photo-thermoelectric, thermionic diode rectification, tunnel diode rectification, and resistive self-mixing (b,c) Calculated dependencies of Seebeck coefficient difference between right and left sections of the device, $S_2 - S_1$, vs respective top gate voltages $V_{\rm tg1}$ and $V_{\rm tg2}$ at $T=25$ K in the gapless state [$V_{\rm bg} - V_{\rm CNP} = 0$ V, (b)] and in the gapped state [$V_{\rm bg} - V_{\rm CNP} = -10$ V, (c)]. (d) Gate-dependent difference of Seebeck coefficients at $T=300$ K (e) Maximum absolute difference between Seebeck coefficients, $\max|S_2 - S_1|$, as a function of band gap $E_G$ at CNP (controlled by the bottom gate) (f,g) Gate-dependent photovoltage due to the non-linearity of the tunnel junctions, $V_{TJ}$, calculated at cryogenic temperatures $T=25$ K and two values of back gate voltage $V_{\rm bg} = - 5$ V (f) and $V_{\rm bg} = - 10$ V (g)}
    \label{fig:theory}
\end{figure}

\textbf{Origins of THz photoresponse.} The most important quantity for determination of microscopic detection mechanism is the sign of generated photovoltage. To this end, we briefly discuss the most widespread detection mechanisms and the anticipated signs of $V_{\rm ph}$ in our $p$-$n$ junction configuration [see the schematic in Fig.~\ref{fig:theory} (a)].

The photo-thermoelectric (PTE) effect arises due to heating of the junction by radiation, and subsequent thermal diffusion of the hot carriers. For both electrons and holes, thermal diffusion occurs from hot junction to cold metal leads  (see figure ~\ref{fig:structures}). As a results, the photovoltage measured at the drain contact is negative. 

The resistive self-mixing effect (RSM) occurs due to the simultaneous action of a vertical THz electric field (which changes the 2d sheet carrier density) and its longitudinal component (which drags the charge carriers). In other words, RSM occurs due to the self-gating nonlinearity of the transistor structure. It is possible to show that RSM in the lateral $p$-$n$ junction results in positive photovoltage at the drain, as opposed to the PTE. Indeed, at one half-period of THz wave, the positive ac potential is induced at the $p$-doped side, and the negative ac potential is induced at the $n$-doped side. Once the top gates are grounded, such ac potential results in enhanced carrier densities in the channel and enhanced conduction. These extra holes are longitudinally dragged from $p$ to $n$-region, while electrons are dragged in opposite direction, leading to positive potential at the drain.

The sign of photovoltage due to $I(V)$ non-linearity of the $p$-$n$ junction depends sensitively on the microscopics of carrier transport. If the transport is governed by thermionic emission, the emerging photocurrent is directed from $p$ to $n$-side (positive photovoltage at the drain). As the positive ac potential difference is induced between source and drain, the diode becomes forward-biased and a large amount of electrons overcomes the barrier from $n$ to $p$ side. If the transport is governed by interband tunneling, the rectified voltage at the drain is, conversely, negative. As the negative ac potential difference is induced between source and drain, the strength junction field goes up leading to large tunnel transparency. This results in elevated current of electrons from the valence band of $p$-region to the conduction band of the $n$-region.

To summarize [see Fig.~\ref{fig:theory} (a)], positive photovoltage in $pn$-configuration is expected for RSM and thermionic diode rectification, while negative is expected for PTE and tunnel-diode mechanisms.

The obtained photovoltage data for cryogenic temperatures can be largely described by PTE rectification scenario. Both measured sign and the number of sign changes appear as expected for thermoelectric mechanism. Within the simplest theoretical model, the PTE photovoltage is $V_{\rm PTE} = (S_2 - S_1)\Delta T_{\rm THz}$, where $S_{1/2}$ are the Seebeck coefficients of BLG sections under left and right gates, and $\Delta T_{\rm THz}$ is the radiation-induced heating of the junction itself. The calculated maps of $S_2(V_1) - S_1(V_2)$, presented in Fig.~\ref{fig:theory} (b-c), accurately represent the sign and sign-changing contours of the experimental responsivity (see Methods for computational details). Further on, it is possible to show that maximum value of Seebeck coefficient difference appears if Fermi level in $n$-region crosses the bottom of conduction band, while that in $p$-region crosses the top of the valence band, in accordance with experimental data. The maximum value of Seebeck coefficient difference, $\max|S_1 - S_2|$, scales linearly with band gap once $E_G \gtrsim k T$ [Fig.~\ref{fig:theory} (e)], which can explain the positive effect of gap on voltage responsivity.  

At the same time, several features of our data do not fit into thermoelectric picture. Indeed, the PTE scenario predicts identical photoresponse for $nn$ and $pp$ junctions, but the experimental data shows strong sensitivity of $r_V$ in these quadrants to the sign of back gate voltage. For $V_{\rm bg} > 0$ (electron doping of near-contact regions), the responsivity is larger in the $pp$-configuration, compared to the $nn$-configuration. It implies that extra junctions at the edges of top gates contribute to the photoresponse, either by providing extra rectification mechanisms, or by changing the flow of hot carriers.

The main candidate mechanism contributing to photovoltage at cryogenic temperatures comes from rectification by tunnel $p$-$n$ junction nonlinearities~\cite{gayduchenko_tunnel_2021}. We have evaluated the respective photovoltage $V_{\rm TJ}$ by constructing a circuit model comprised of non-linear junctions at the boundaries of regions with dissimilar doping, and sequential bulk sections with linear current-voltage characteristics. The result of such modeling is shown in Fig.~\ref{fig:theory} (f,g). Overlay of basic thermoelectric response with additional tunnel response provides the desired effect: the absolute value of photovoltage is enhanced if the doping under top gates is opposite to that in the near-contact sections.

At the room temperature the dominant rectification mechanism changes, since the photovoltage sign is reversed. The two remaining candidate mechanisms providing the necessary sign are resistive self-mixing and thermionic non-linearity of the tunnel junction. The latter is unlikely, as the induced gap $E_G \sim 20$~meV is well below $2kT$ at room temperature. Thus, the thermionic $I(V)$ characteristic should be almost linear in our experiment. Strong suppression of thermoelectric voltage at room temperature can be attributed to fast energy relaxation of heated electrons via emission of optical phonons. At the same time, the resistive self-mixing is relatively robust to temperature variations, and its magnitude depends only on FET transconductance, $V_{RSM} \propto d \ln G/dV_g$~\cite{bandurin_dual_2018}.

\begin{figure}[ht!]
    \centering
    \includegraphics[width=1\textwidth]{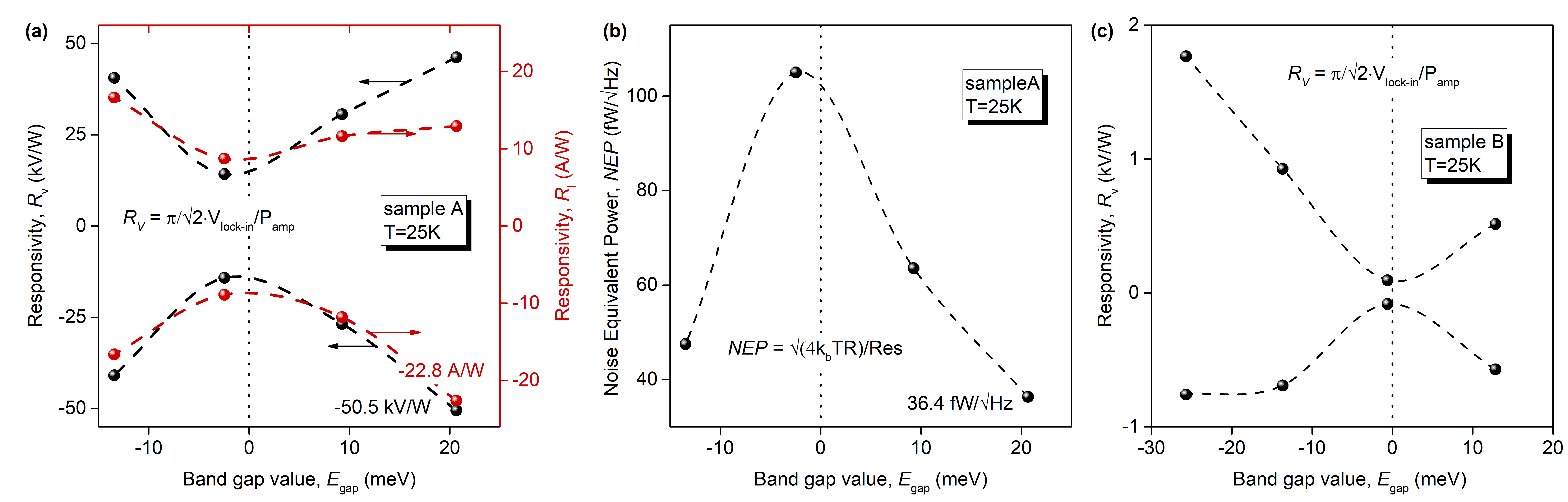}
    \caption{{\bf Practical figures of merit for BLG-based $p$-$n$ junction detectors.} (a) Maximum voltage and current responsivities for sample A vs band gap (b)  Minimum attainable NEP for sample A vs band gap (c) Maximum voltage responsivity for sample B vs band gap. Dashed lines in (a,b,c) serve as guides for the eye }
    \label{fig:bandgap}
\end{figure}

\textbf{Practical figures of merit.} Induction of band gap results in considerable enhancement of photovoltage in our detectors. This leads, in turn, to superior practical figures of merit, responsivity and noise equivalent power. The maximum attainable values of external responsivity are as large as 50 kV/W for sample A and 1.8~kV/W for sample B [Fig.~\ref{fig:bandgap} (a,c)]. At the same time, the enhancement factor of voltage resposivity, defined as the ratio of signals at maximum $E_G$ and at $E_G=0$, is larger for sample B and reaches $\approx 19$, while for sample A it reaches $\approx 3.3$. Instructively, the maximum current responsivity of both devices is also a growing function of the band gap, with the ultimate value as large as 20~A/W for sample A.

Another important figure of merit telling about minimum electromagnetic signal detectable in the noisy background is the noise equivalent power NEP. It is defined as the ratio of voltage noise spectral density $s_V$ and voltage responsivity $r_V$, ${\rm NEP} = s_V/r_V$. The zero-bias operation of $p$-$n$ junction detectors ensures that the only source of noise comes from thermal voltage fluctuations, thus $s_V = \sqrt{4 kT R_{SD}}$. Direct evaluation shows that NEP becomes also becomes smaller (better) upon band gap induction [Fig.~\ref{fig:bandgap} (b)]. It falls from $\sim 100$~fW/Hz$^{1/2}$ in the gapless state of sample A, down to 36~fW/Hz$^{1/2}$ at the maximum band gap. The improvements in NEP may look intriguing taking into account the exponential growth of resistance at CNP $R_{\rm CNP} \propto \exp\{E_G/2kT\}$, which should seemingly increase the noise. However, the photoresponse in gapped BLG is maximized when the Fermi level is located in the vicinity of band edges. In such a situation, the resistance and noise are weakly sensitive to the gap opening.

The minimum attained NEP of 36~fW/Hz$^{1/2}$ is about five times lower than that reported previously for cryogenically cooled BLG-based field-effect transistors ($\approx 200$~fW/Hz$^{1/2}$ \cite{gayduchenko_tunnel_2021}). A probable origin for supremacy of our device, compared to the previously reported one, is the scheme of antenna coupling between source and drain (not between source and gate). In the previous schemes, the photovoltages emerging at source and drain junctions are competing, while in the current scheme the major part of voltage builds up at an uncompensated centran $p$-$n$ junction. Our achieved NEP is also lower than that reported for BLG-based bolometers~\cite{yan_dual-gated_2012}, considering the fact that internal responsivity was reported in Ref.~\cite{yan_dual-gated_2012}. In general, the reported values of cryogenic NEP are highly competitive with those for superconducting~\cite{supercond1,supercond2} and semiconducting~\cite{irlabs} bolometers available on market. 

The room temperature NEP in our detector falls down to 160~pW/Hz$^{1/2}$. This value is two  times larger than that reported for single graphene-layer $p$-$n$ junction~\cite{castilla_fast_2019} and 6 times larger than that for state-of-the art GaN-based transistors~\cite{Bauer2019}; the probable origin is that we did not intentionally optimize the matching of our detector to the antenna. It is also $\sim 5$ times smaller than that reported for GaAs quantum well-based sub-THz detectors~\cite{Shchepetilnikov2019}

\textbf{Conclusions.} We have revealed two possible scenarios of responsivity enhancement with band gap induction in BLG $p$-$n$-junctions that are consistent with our data, larger Seebeck coefficient and stronger tunneling rectification. Still, several aspects of data require deeper theoretical insights. In particular, simultaneous enhancement of current and voltage responsivities with $E_G$ cannot be explained by variations of Seebeck coefficient $S$ solely. Indeed, assuming the voltage responsivity to scale as $r_{V,\rm max}\propto S_{\rm max} \propto 1/E_G$, and accounting for conductivity reduction with $E_G$, we see that current responsivity $r_I$ should remain constant. One has thus to consider other mechanisms of responsivity enhancement, beyond variations of $S$. As example, the thermal conductivity of electrons $\chi_e$ in BLG can be also suppressed by finite gap. This refers both to the bulk value of $\chi$ and finite thermal resistivity of the $p$-$n$-junctions. These factors would result in higher carrier temperatures $\delta T_{\rm THz}$. More thorough studies of functional dependence $r_V(E_g)$ in a broader range of gaps may shed light on these questions.

Bilayer graphene was selected only as a convenient platform to study the effect of band gap (being tunable) on $p$-$n$ junctions photoresponse. The conclusion about positive effect of gap on thermoelectric and tunneling rectification is, however, quite generic. It should apply to other 2d systems with moderate gap $\sim 0.1$~eV, such as black phosphorous~\cite{Black_P_THz}, palladium diselenide~\cite{PdSe2_THz}, and mercury cadmium telluride quantum wells~\cite{CdHgTe_THz}. As a limitation, very large band gaps can lead to large $p$-$n$ junction resistance both via suppression of tunneling and thermionic processes, mismatch with antennas and increased thermal noise.

In spite of the remaining physics challenges, the revealed dependences of photoresponse on induced band gap in BLG junctions demonstrate that there's a plenty of space for further sensitivity optimization. The gaps induced in our work $E_G \lesssim 25$ meV are well below the maximum values of $250$~meV reported from optical experiments~\cite{zhang_direct_2009} and $100$~meV reported in dc measurements of temperature-dependent resistance~\cite{icking_transport_2022}. As a result, we can expect further four- to ten-fold responsivity enhancement in maximally gapped devices. This should lead to NEP as low as $4...9$ fW/Hz$^{1/2}$ at cryogenic temperatures. Room-temperature operation of maximally gapped devices should also lead to reduced noise-equivalent powers, which, already under current conditions, is comparable with existing detectors. 

\section*{Methods}

\textbf{Fabrication.} The devices were fabricated by mechanical exfoliation of graphite bulk crystals onto a 280~nm SiO$_2$/Si wafer. Heterostructures hBN-BLG-hBN was stacked and then placed on top of SiO$_2$/Si wafer using standard dry transfer technique described elsewhere ~\cite{purdie_cleaning_2018}.
The channel length, Ti/Au 1D metal contacts to graphene \cite{wang_one-dimensional_2013} and top gates were defined by electron beam lithography (EBL) and dry etching in SF$_6$ plasma (see fabrication details in Supplementary Materials).

\textbf{Measurements} The sample was placed in an optical cryostat with temperature tunable in the range from 25 to 300~K. The cryostat window was made of thin ($\approx 80\mu$m) THz-transparent mylar film. The radiation was linearly polarized along the antenna axis. We used a set of filters to reduce the radiation intensity (up to $3\times 10^{-4}$ of the initial power) so that the graphene detector operated in a linear mode and did not demonstrate saturation. 
We made strong precautions to illuminate the sample in a symmetrical way. To this end, we measured the photovoltage maps as a function of $x$-$y$ coordinates. We made sure that at opposite channel doping ($pn$ and $np$) the signal changes sign, as it should be from symmetry considerations.

We used two two-channel Source-Meters Keithley SM2612B and SM2636B 
to independently apply dc voltages to each of the three gates and simultaneously check the leakage current. To simultaneously measure resistance, photovoltage and photoresistance, we used a standard lock-in technique with several Lock-in Amplifiers SR860. To measure the BLG resistance we applied small ac current (200~nA rms) via  lock-in at 111~Hz frequency, while THz source was modulated at 33~Hz frequency. Electrical signals at non-multiple frequencies do not mix, so we measured the net photovoltage on the unbiased channel.

\textbf{THz source calibration and evaluation of responsivity.} We calibrated our THz source (IMPATT diode) using a pyroelectric detector.
First of all, we checked that the output power of our source is independent of its modulation frequency (which is tunable from 11~Hz to 39~Hz). Then, we compared our 130~GHz source with 140~GHz reference source by pyroelectric detector \#1 at the same modulation frequency. Third, the signal from pyroelectric detector \#1 with frequency-dependent absorption was normalized by pyroelectric detector \#2 with metallic absorber and flat response over the frequency to extract the correction coefficient between absorption at 130~GHz and 140~GHz. Finally 140~GHz source in continuous wave regime was measured on VDI Erickson PM5B Power Meter connected by waveguide port, and thus we was able to recalculate the measured integrated power from pyroelectric detector \#1 to peak-to-peak power of our source in mW.

We also measured the output radiation beam profile using a motorized xyz-stage, and obtained the Gaussian beam with a divergence angle of $\sigma=10^\circ$. The output peak power of our detector appears to be $P=16.4$~mW. The power absorbed by the sample depends on the parameters of the optical setup. Overall, the total power reached the samples A and B was $P_{\rm 0 ,A} = 18$~nW and $P_{\rm 0, B} = 340$~nW. More details can be found in Supplementary Materials.

External responsivity is the relation between photovoltage and power reaching the device $r_{\rm V}=V_{\rm ph}/P_0$. The photovoltage follows the intensity of modulated light instantaneously, and its time dependence represents a series of rectangular pulses with 50\% filling factor and height of $V_{\rm ph}$. The ratio between $V_{\rm ph}$ and lock-in signal $V_{\rm ph,  lock-in}$ is   $V_{\rm ph} = \pi/2 \cdot \sqrt{2} \cdot V_{\rm ph, lock-in}$. The factor of $\pi/2$ comes from Fourier expansion of rectangular pulses, and factor of $\sqrt{2}$ -- from converting rms voltage to peak-to-peak voltage. 

\textbf{Modeling of detector response.} To estimate the thermoelectric photovoltage of BLG detector, we use the relation $V_{\rm PTE} = (S_2 - S_1)\Delta T_{\rm THz}$. Here $S_2$ and $S_1$ are the Seebeck coefficients of regions under two top gates, and $\Delta T_{\rm THz}$ is the THz-induced heating of the junction. The Seebeck coefficient is obtained from kinetic coefficients for electrons and holes $\alpha_{e/h}$ and $\sigma_{e/h}$, where $\alpha$ relates charge current and thermal gradient ${\bf j} = \alpha \nabla T$, and $\sigma$ is the ordinary electric conductivity:
\begin{equation}
    S = \frac{\alpha_e - \alpha_h}{\sigma_e + \sigma_h}.
\end{equation}
The transport coefficients are expressed from Boltzmann kinetic theory:
\begin{gather}
    \alpha_e (\mu) = \frac{e}{2k} \int\limits_{E_C}^{\infty}{dE \rho(E) v^2(E) \tau_p(E) \frac{E - \mu}{k T} \frac{\partial f_0 (E,\mu)}{\partial E}}, \\ 
    \sigma_e (\mu) = - \frac{e^2}{2} \int\limits_{E_C}^{\infty}{dE \rho(E) v^2(E) \tau_p(E) \frac{\partial f_0(E,\mu)}{\partial E}}.
\end{gather}
Above, $\rho(E)$ is the density of states, $\tau_p(E)$ is the momentum relaxation time, $v(E)$ is the modulus of band velocity, and $f_0$ is the equilibrium Fermi function. The transport coefficients for holes can be found from $\alpha_h(\mu) = \alpha_e(-\mu)$, $\sigma_h(\mu) = \sigma_e(-\mu)$. For further calculations, we assume $\rho(E) = {\rm const}$, $\tau_{p}(E) = {\rm const}$, the resulting dependence of $S$ on band gap weakly depends on these assumptions. The band gap  at given gate voltages was determined using the self-consistent scheme described in~\cite{Falko_bandgap_theory}.

Evaluation of photovoltage due to tunnel rectification $V_{\rm TJ}$ was based on an equivalent circuit comprised of six rectifying elements: two double-gated regions surrounded by four tunnel junctions. We assume the single-gated regions are short enough to be neglected. The band diagram of the transistor was obtained by self-consistently solving the electrostatic equations for bilayer graphene, as described elsewhere~\cite{gayduchenko_tunnel_2021}. Current-voltage characteristics of the tunnel junctions were calculated using the quasiclassical approximation for the barrier transparency~\cite{gayduchenko_tunnel_2021}. The response of double-gated regions to ac perturbations was calculated using the Drude model to take into account the kinetic inductance of electrons in bilayer graphene (eq. S28 in the Supplementary Information of Ref.~\citenum{gayduchenko_tunnel_2021}). Knowing the ac and dc current-voltage characteristics of all the rectifying elements, we obtain the responsivity of the whole transistor by expanding all voltages and currents up to the second order in the input voltage (provided by the antenna) and employing Kirchhoff's laws.

\subsection*{Conflict of interest}
The authors have no conflicts of interest to disclose.

\section{Acknowledgment}
This work was supported by grant \# 21-79-20225 of the Russian Science Foundation (device fabrication, electrical and THz measurements, data analysis and modeling).  The devices were fabricated using the equipment of the Center of Shared Research Facilities (MIPT). S.Z. acknowledges the support of the the Ministry of Science and Higher Education of the Russian Federation (grant \# FSMG-2021-0005). E.T., M.K., and K.N. thank Vladimir Potanin via Brain and Conscionsness Research Center.

\bibliography{Refs}

\end{document}


\maketitle

\newpage

\section{I. Device fabrication}
\begin{figure}
    \centering
    \includegraphics[width=1\textwidth]{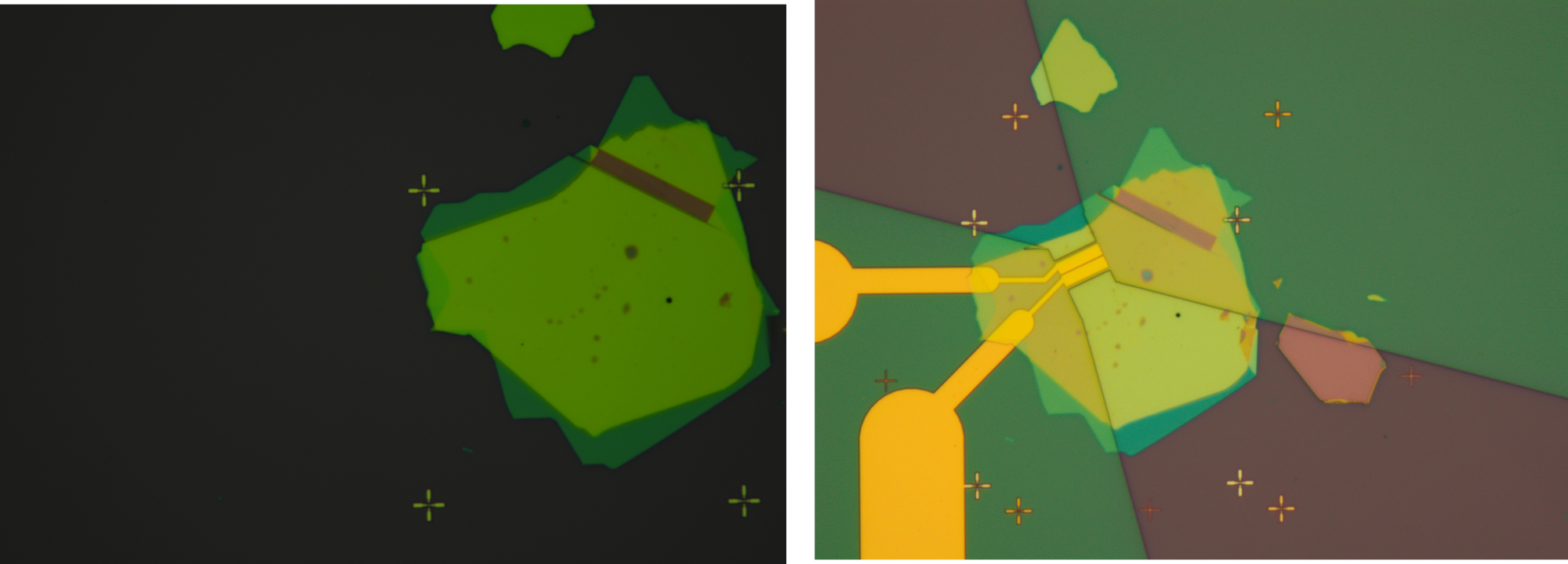}
    \caption{hBN-BLG-hBN stack and lithography-defined contacts of the sample A}
    \label{fig:}
\end{figure}

\begin{figure}
    \centering
    \includegraphics[width=1\textwidth]{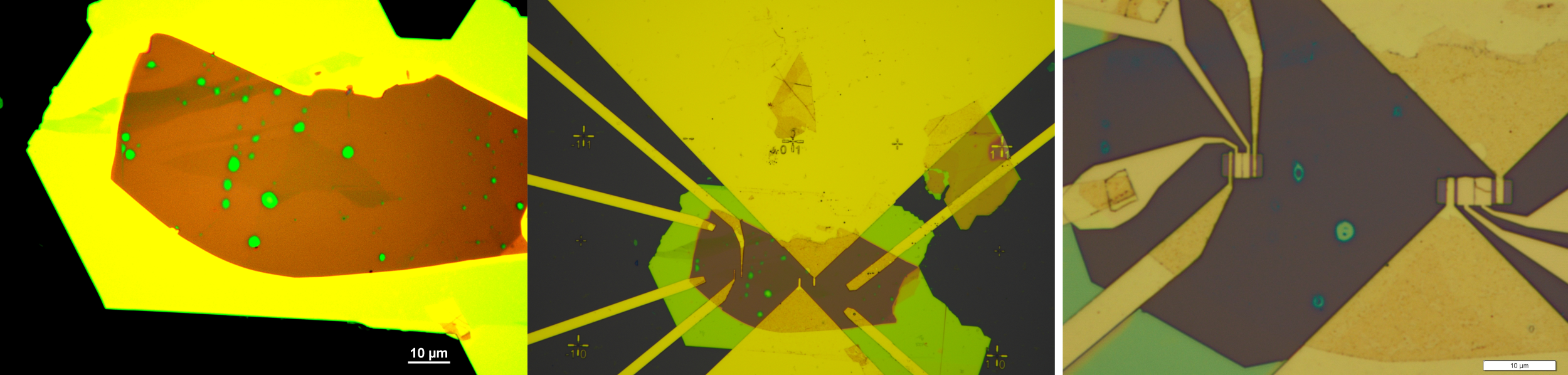}
    \caption{hBN-BLG-hBN stack, metal contacts and final view of the sample B (we measured the right-placed device)}
    \label{fig:}
\end{figure}

BLG flakes were identified after exfoliation by optical contrast using an optical microscope. Hexagonal boron nitride (hBN) flakes about 50nm thick were exfoliated from hBN bulk crystals and found in the same way as graphene flakes. 
We used e-beam lithography with PMMA A4, developing in cold IPA:DI water 3:1 and DI water as a stopper. After developing we etched the sample in oxygen plasma of 150W for 2-3s to remove residual resist in developed windows. To define the channel shapes we etched hBN-graphene outside of the future transistor channel in SF6 plasma. Ti/Au metal contacts was deposited by electron beam sputtering followed by a lift-off process in acetone. We did three rounds of electron lithography to fabricate the device - the first to make the top gates, the second to sputter the drain-source contacts, and the third to etch the excess hBN and graphene from the sample.

\section{II. THz source calibration}

\begin{figure}
    \centering
    \includegraphics[width=1\textwidth]{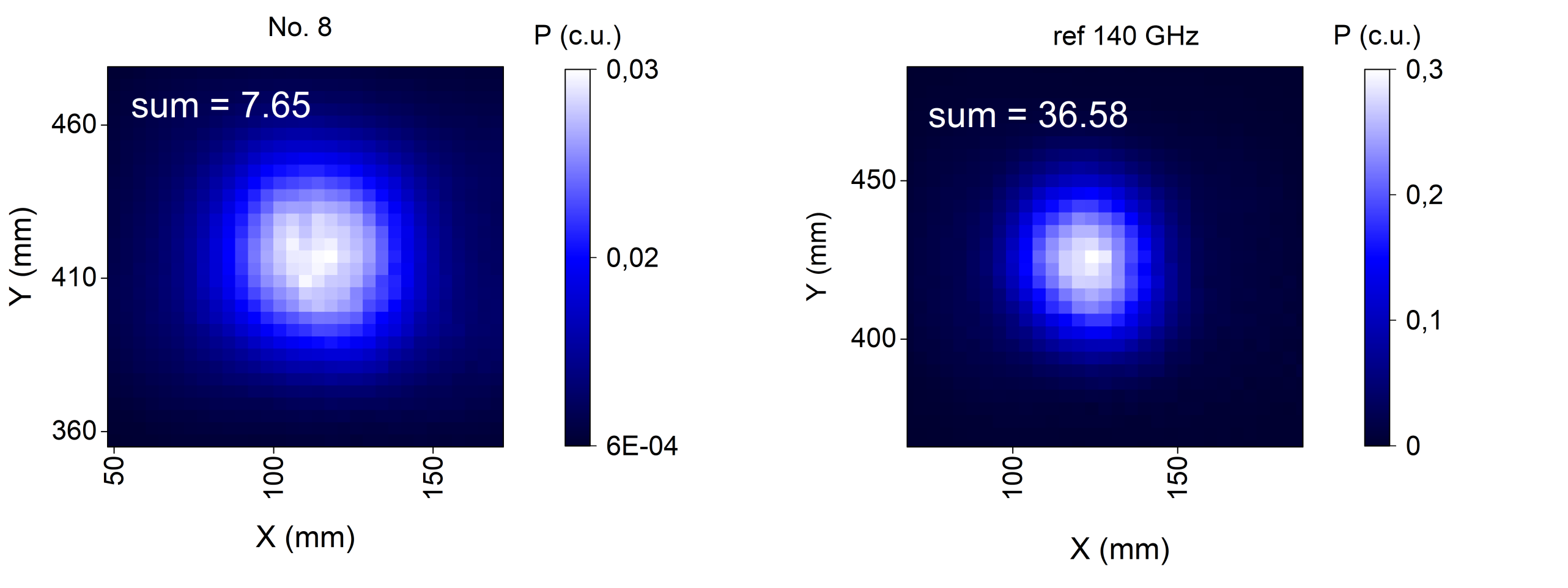}
    \caption{Integrated power of our 130 GHz source and 140 GHz reference source at a modulation frequency of 23 Hz, scanned by a \#1 pyroelectric detector mounted on a motorized x-y stage.}
    \label{fig:}
\end{figure}

\begin{figure}
    \centering
    \includegraphics[width=1\textwidth]{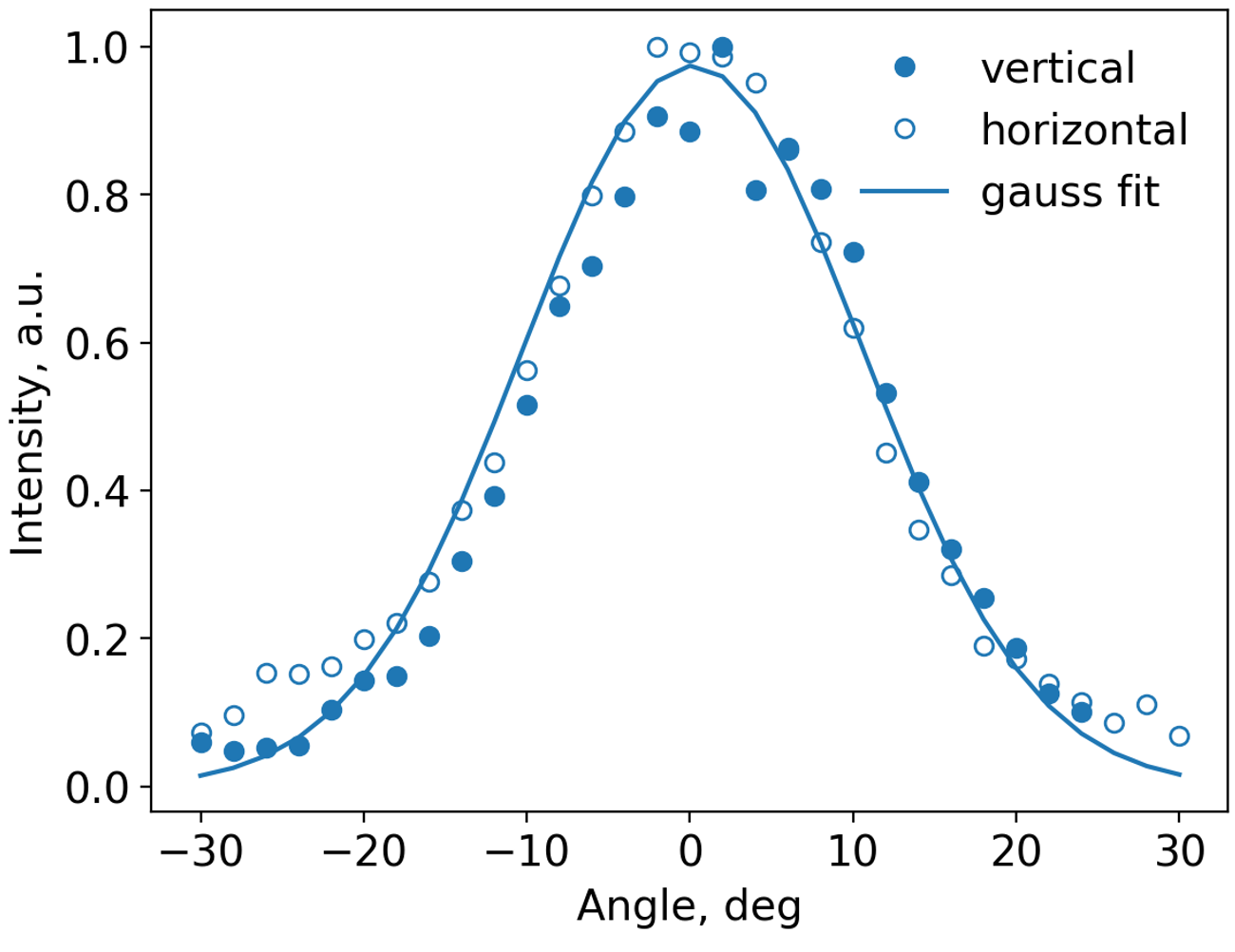}
    \caption{Measured THz source beam angle distribution in vertical and horizontal planes along with a Gaussian fit with $\sigma=10^o$.}
    \label{fig:}
\end{figure}

\section{III. Measurements}
The pressure inside the cryostat was maintained no more than 1e-5hPa. We used 0.13 THz IMPATT-diode source with horn antenna output.
We characterized sample A in a completely symmetrical configuration, applying $-I_{sd}/2$ and $+I_{sd}/2$ to the source and drain, respectively, and measured all parameters ($R_{\rm s-d}, V_{\rm ph}$ and $\delta R$) simultaneously as differential signals $V_d - V_s$. Sample B was characterized with grounded source, and we measured in series the resistance and then the photovoltage from this sample.
Sample A was measured without Si lens at a distance of about 19 cm from the THz source, using filter with transmittance 0.03, and polarizer at $45^o$ having the transmittance 0.5. We considered the sample absorption cross section to be equal to the maximum possible, resting on the fundamental limit \footnote{Mylnikov, D.; Svintsov, D. Physical Review Applied 2022, 17, 064055.} $3/2\cdot \lambda^2/4\pi \approx 0.6mm^2$, so we present here the values of photoresponsivity and equivalent noise power estimated from below.
Sample B was measured with Si lens 12mm in diameter, and the sample was located at a distance of 17.5 cm from the THz source, filter 0.003, and the same polarizer angle, set parallel to the antenna. Assuming the lens focuses all radiation into a spot smaller than antenna ($1mm^2$), we also considered the reflectance of external radiation from the lens surface (26\%).
The total power reached the sample B is $P_{\rm 0} = 340 nW$

\begin{figure}
    \centering
    \includegraphics[width=1\textwidth]{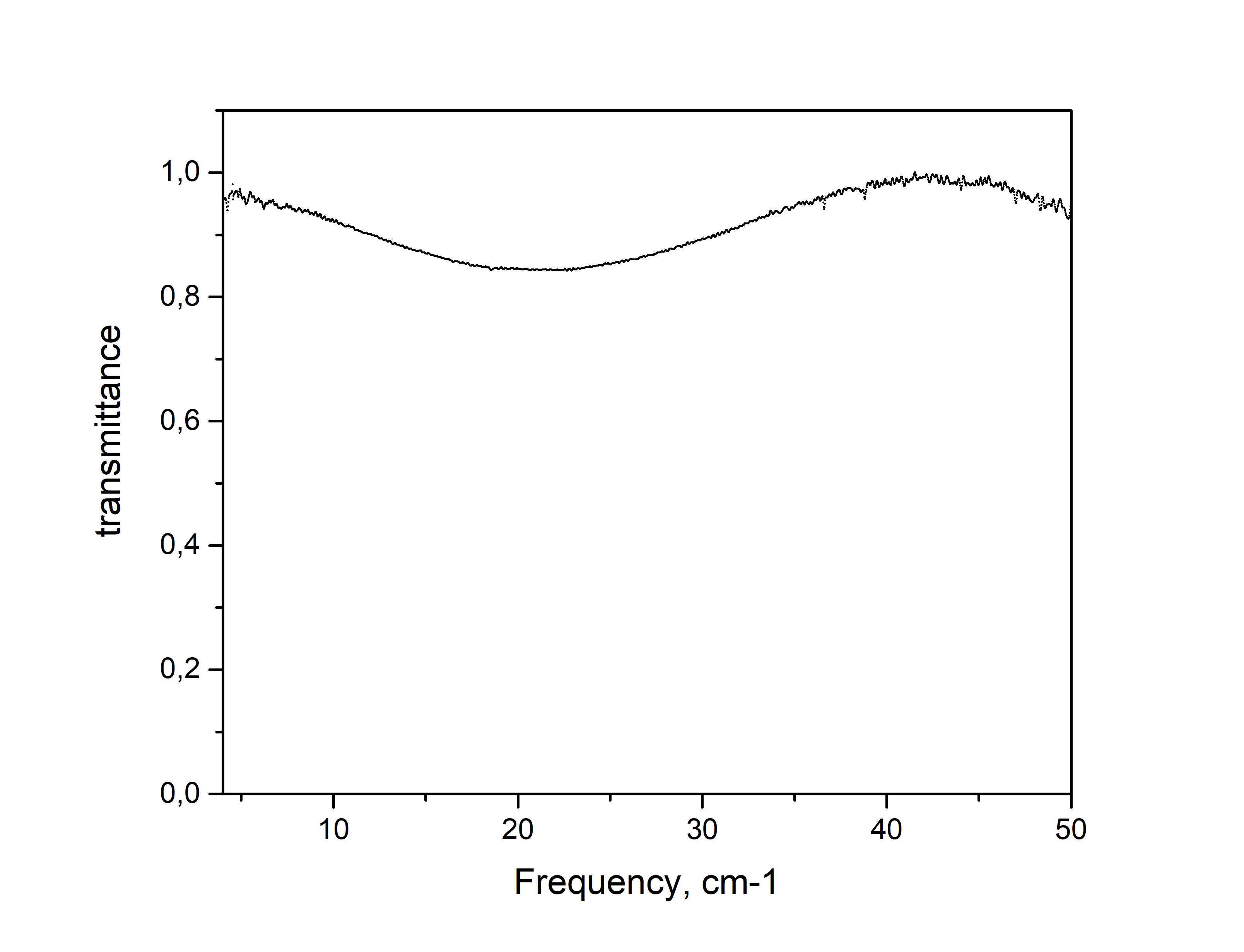}
    \caption{Transmittance of window material for cryostat - mylar film. It has an almost constant transmittance, about 100\%.}
    \label{fig:}
\end{figure}

\section{IV.Additional results}
\begin{figure}[H]
    \centering
    \includegraphics[width=1\textwidth]{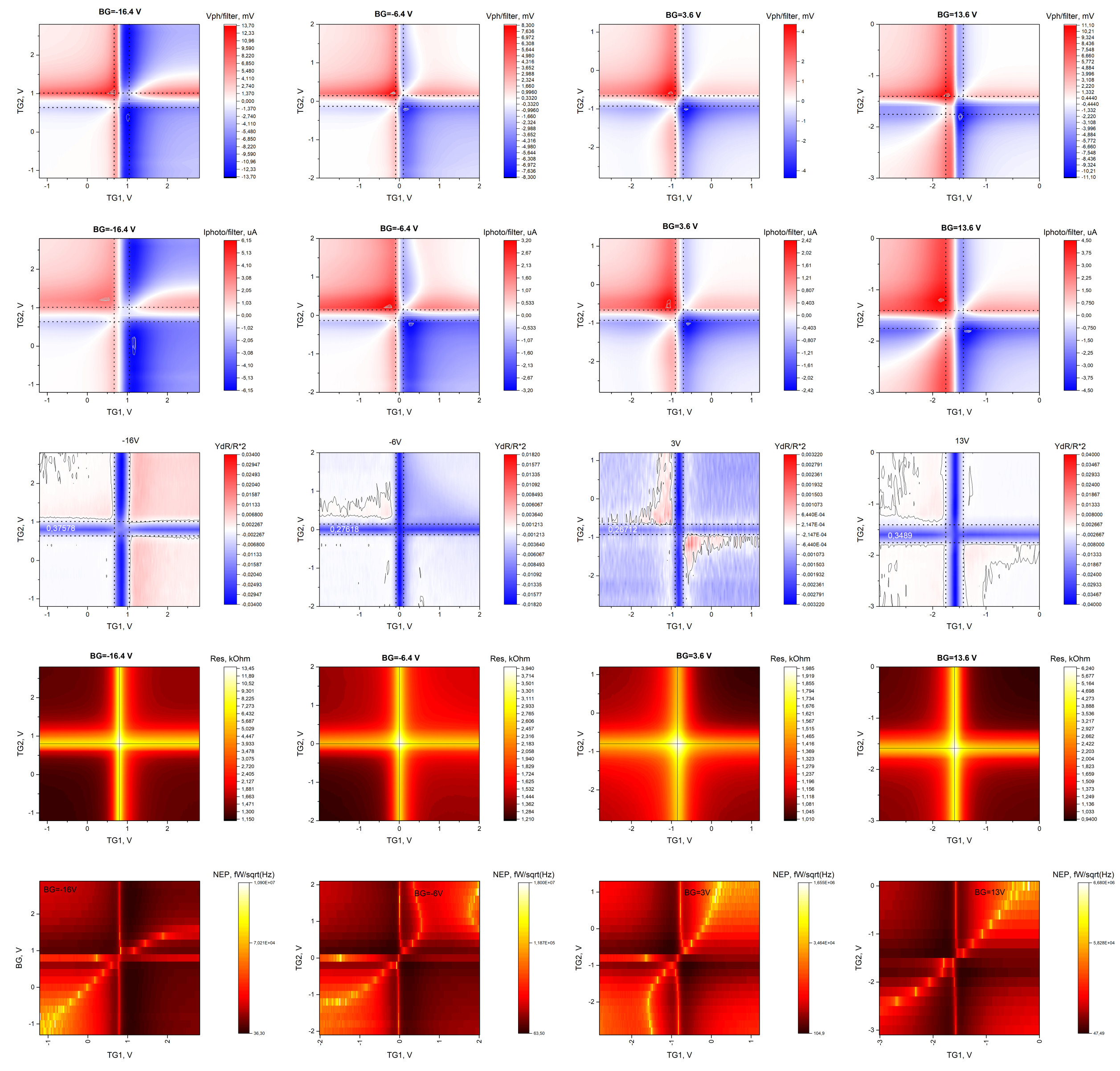}
    \caption{Photovoltage, photocurrent, photoresistance, source-to-drain resistance and NEP for the sample A at different fixed back gate voltages at 25K.}
    \label{fig:}
\end{figure}

\begin{figure}[H]
    \centering
    \includegraphics[width=1\textwidth]{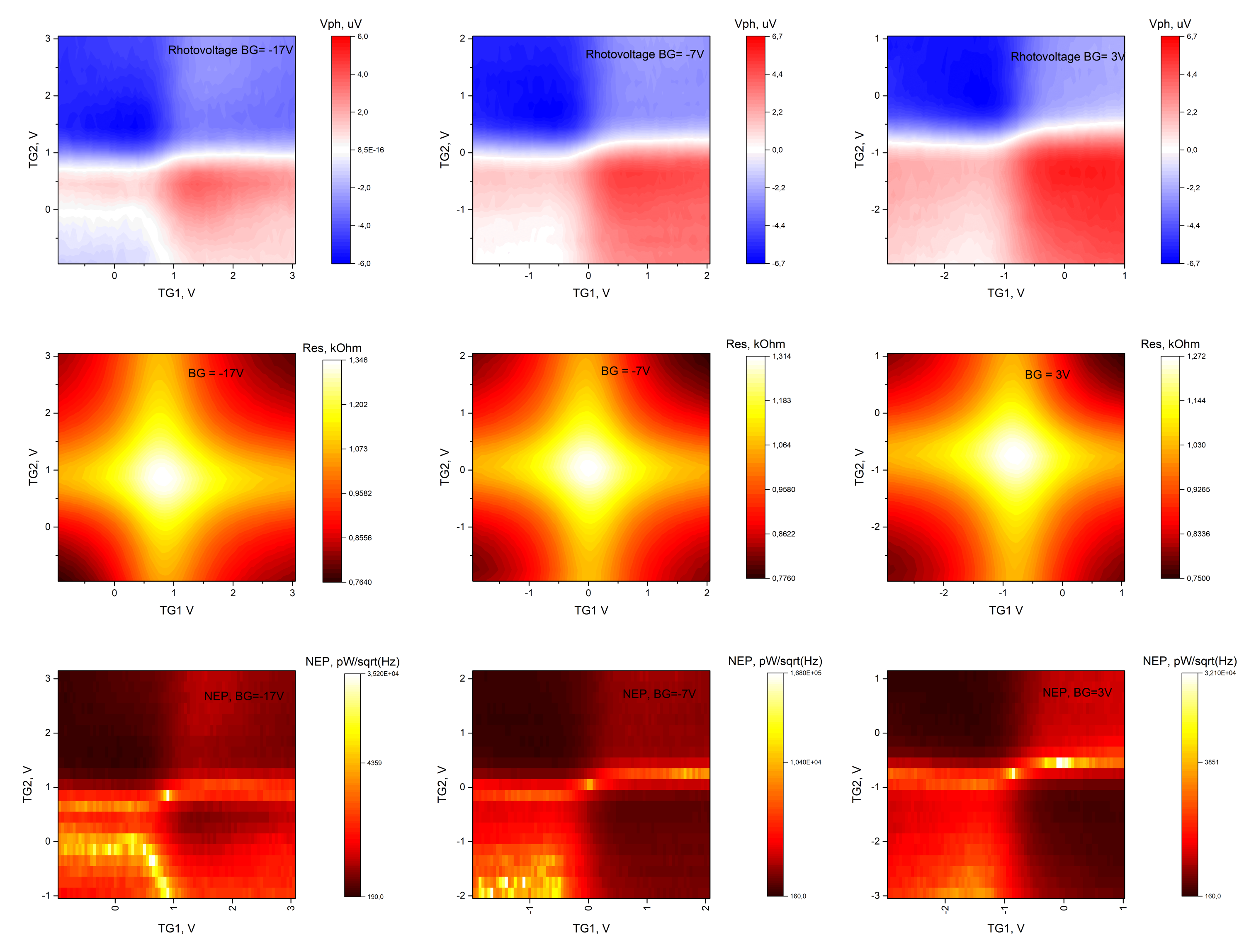}
    \caption{ Photovoltage, source-to-drain resistance and NEP for the sample A at different fixed back gate voltages at 300K.}
    \label{fig:}
\end{figure}

\begin{figure}[H]
    \centering
    \includegraphics[width=1\textwidth]{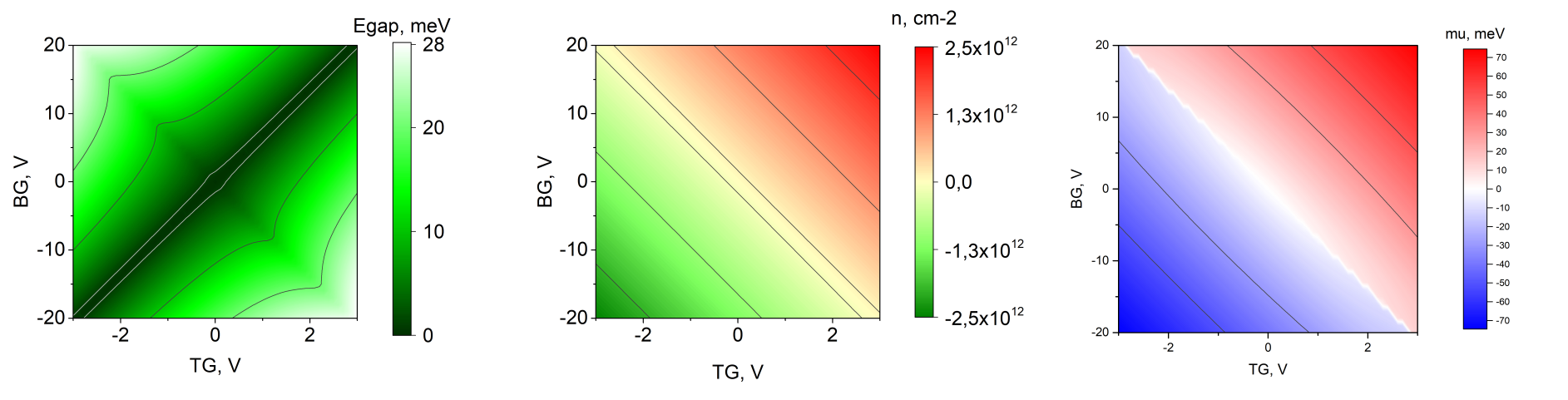}
    \caption{Calculated values of band gap, concentration and chemical potential as a function of top and back gate voltages}
    \label{fig:}
\end{figure}


\begin{figure}[H]
    \centering
    \includegraphics[width=1\textwidth]{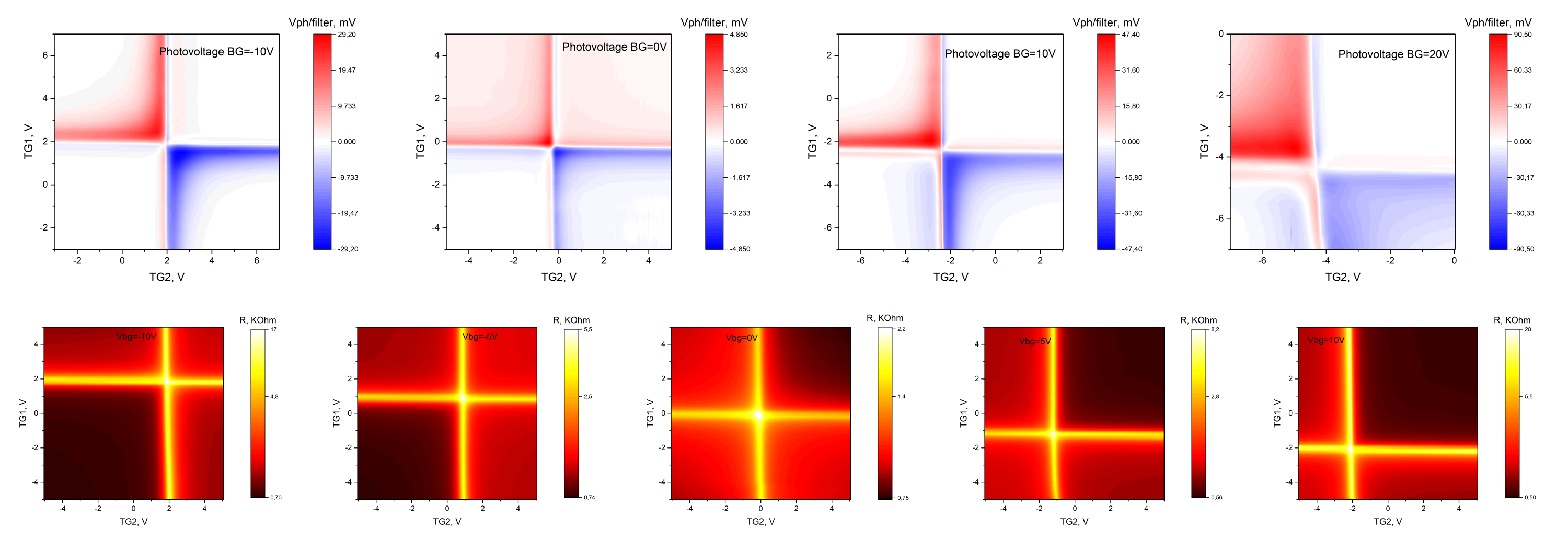}
    \caption{Photovoltage and source-to-drain resistance for the sample B at different fixed back gate voltages at 25K.}
    \label{fig:}
\end{figure}
